# Securing the Internet of Things in the Age of Machine Learning and Software-defined Networking

Francesco Restuccia, *Member, IEEE,* Salvatore D'Oro, *Member, IEEE,* and Tommaso Melodia, *Fellow, IEEE*

*Abstract*—The Internet of Things (IoT) realizes a vision where billions of interconnected devices are deployed just about everywhere, from inside our bodies to the most remote areas of the globe. As the IoT will soon pervade every aspect of our lives and will be accessible from anywhere, addressing critical IoT security threats is now more important than ever. Traditional approaches where security is applied as an afterthought and as a "patch" against known attacks are insufficient. Indeed, next-generation IoT challenges will require a new secure-by-design vision, where threats are addressed proactively and IoT devices learn to dynamically adapt to different threats. To this end, machine learning and software-defined networking will be key to provide both reconfigurability and intelligence to the IoT devices. In this paper, we first provide a taxonomy and survey the state of the art in IoT security research, and offer a roadmap of concrete research challenges related to the application of machine learning and software-defined networking to address existing and next-generation IoT security threats.

*Index Terms*—Internet of Things, Security, Trust, Authentication, Survey, Challenges, Perspective

## I. Introduction

Imagine a world where our coffee machine communicates with our alarm clocks to produce perfectly brewed coffee in the morning, and tells the bedroom curtains when to open based on our past preferences; where our cars automatically melt the ice accumulated in an overnight snow storm, our refrigerators asks an online store to deliver us a number of missing groceries through drones, and our doctors prescribe us medicines using biometrical data collected from tiny sensors implanted inside our bodies. Although 20 years ago this world might as well have been the product of a talented sci-fi writer, recent advances in the fields of computer science and engineering are today allowing the dream of the *Internet of Things* (IoT) to become reality.

Cisco Systems has forecast that by 2020, over 50 billion connected "things" will be absorbed into the Internet, including cars, kitchen appliances, televisions, surveillance cameras, smartphones, utility meters, intra-body sensors, thermostats, and almost anything we can imagine [1]. Accordingly, it has been predicted that annual revenues could exceed $470B for the IoT vendors selling the hardware, software and comprehensive solutions for the IoT [2]. The application of IoT products and services will pervade every sector and industry from smart home and smart city, education, health-care, manufacturing, mining, utilities, commerce, transportation, surveillance, infrastructure management, to supply chain and logistics. The opportunities presented by the IoT are endless, and its full potential will soon be tangible since more and more devices are getting connected to the Internet every day.

While the benefits of IoT are undeniable, the reality is that security is not keeping up with the pace of innovation. As the IoT will pervasively expand, it is expected that its heterogeneity and scale will magnify existing Internet security threats. Once humans, sensors, cars, robots, and drones are able to seamlessly interact with each other from any side of the globe through the IoT, a number of threats that we cannot even imagine today will be unveiled. If necessary precautions are not taken, malicious individuals will leverage the pervasiveness of the IoT to disrupt communications, gain significant monetary advantages, or even physically harm people. For example, researchers have found critical vulnerabilities in a wide range of IoT baby monitors [3], which could be leveraged by hackers to carry out a number of nefarious activities, including authorizing other users to remotely view and control the monitor. In another development, it was proven that Internet-connected cars may be remotely controlled [4], including operations such as unlocking the doors or even shutting down the car in motion. Some of the most worrisome cases of IoT hacks, however, involve medical devices and can have fatal consequences on patients' health [5, 6].

In other words, in a few years *we will need to entrust the IoT network with our own lives*. Thus, addressing the security and privacy issues of the IoT has now become more important than ever. Realizing the importance of solving this spinous problem, the U.S. Senate has very recently proposed the bipartisan Internet of Things Cybersecurity Improvements Act of 2017 [7], which seeks to drive security in Internet-connected devices. The proposed legislation requires vendor commitments to ensure that devices do not contain known security vulnerabilities when shipped, to ensure proper disclosure of new security vulnerabilities, and to prepare remediation plans for any IoT device where known vulnerabilities have been discovered. This implies that IoT manufacturers will need to be *proactive and reactive* as far as security is concerned. Similar efforts have been put forth by the Department of Homeland Security (DHS), which has recently outlined the department's strategic principles for securing the IoT in [8]. The DHS's memorandum explicitly mentions that "security should be evaluated as an integral component of any network-connected device," and that IoT security should be "design[ed] with system and operational disruption in mind." Recognizing the need for a secure and highly dependable IoT infrastructure, the National Science Foundation (NSF) has formulated the Future Internet Architecture program to stimulate innovative and creative research to explore, design, and evaluate trustworthy future Internet architectures [9].

F. Restuccia, S. D'Oro and T. Melodia are with the Department of Electrical and Computer Engineering, Northeastern University, Boston, MA, 02115 USA e-mail: {frestuc, salvatoredoro, melodia}@ece.neu.edu.

Perhaps the biggest challenge in securing the IoT is that a plethora of *heterogeneous technologies, protocols, and requirements will necessarily co-exist*. As a consequence, a security measure that is appropriate for one IoT device may not be appropriate for another, and different devices will have different requirements in terms of security levels and objectives. Furthermore, it is still unclear who will be responsible for security decisions in a world where one company designs a device, another supplies component software, another operates the network in which the device is embedded, and another deploys the IoT device. Even if all the components were developed by a single manufacturer, similar IoT devices would be manufactured by several different companies worldwide, which will necessarily attempt to impose themselves in their specific market sector by customizing and isolating their protocols and technologies.

Another crucial aspect is that most of the existing security countermeasures are based on computation-expensive, high-overhead algorithms and protocols [10]. However, *the stringent budget constraints impose limited memory and computational power* in IoT platforms, as well as the employment of tiny, inexpensive batteries for energy storage. This implies that IoT security technologies must accommodate the constraints of IoT devices. It is also important to point out that due to the distributed nature of the IoT, most of the generated traffic will be wireless in nature [11]. On the other hand, wireless networks are extremely vulnerable to a plethora of security threats, including eavesdropping, denial-of-service (DoS), spoofing, message falsification/injection, and jamming, just to name a few [12], which are usually hard to predict and significantly dynamic in nature [13]. These attacks are relatively simple to carry out, largely because existing IoT platforms are *inflexible in terms of both software and hardware architecture*. Thus, they are not able to withstand dynamic and complex security attacks conducted through the wireless medium. On the other hand, to tackle existing and future security threats in the IoT, *adaptation and self-healing will play a key role*, as the next-generation IoT must be able to face unexpected changes of the environment [10, 14].

<u>*Novel Contributions.*</u> Our vision is simple: the challenges mentioned above cannot be addressed without a radically different approach to the design of secure IoT systems. For this reason, in this paper we advocate the need for a different approach to the design of **secure-by-design** IoT systems, where threats are detected through **learning** and mitigated through **polymorphic** software and hardware architectures.

We have highlighted in bold a keyword that epitomizes an important aspect of next-generation secure IoT systems. Below, we briefly discuss each keyword and introduce the related research challenges that will be discussed in this paper.

- *Security-by-design.* We believe that it is time to enact a *holistic, cradle-to-grave* approach to IoT security, where the system is built to be as free of vulnerabilities as possible, through measures such as continuous testing, authentication safeguards and adherence to best practices. This is because addressing existing vulnerabilities and patching security holes as they are found can be a hit-and-miss process, and will never be as effective as designing systems to be as secure as possible from the beginning. Indeed, building IoT security at the design phase reduces potential disruptions and avoids the much more difficult and expensive endeavor of attempting to add security to products after they have been deployed.
- *Learning.* Traditional security countermeasures are usually tailored to address specific threats under specific network circumstances. However, malicious activity is dynamic in nature and every threat cannot be addressed beforehand. One example are *cross-layer* attacks [15–17], which have *activities* and *objectives* that entail different layers of the network protocol stack. In cross-layer attacks, the adversary may choose to attack a different layer instead of attacking the target layer directly. In this way, small-scale (and thus, hard-to-detect) attack activities may lead to dramatic changes on the target layer. To address these threats, machine learning looks extremely promising as a way to develop security systems that learn to detect and mitigate dynamic attacks effectively.
- *Polymorphic.* Next-generation IoT security will need to be deeply integrated in both hardware and application software layers. Furthermore, bolted-on security mechanisms will not be able to withstand dynamic attacks at multiple level of the network stack. For this reason, software-defined networking could be used as a tool to create technical designs that implement *polymorphic, context-aware* security measures, able to sense and respond to a range of attacks by changing the hardware and software structures of IoT devices to respond to security threats.

In this paper, we provide a taxonomy of existing IoT security threats, and offer to the research community a roadmap of novel and exciting research challenges on applying machine learning and software-defined networking concepts to address IoT security threats. We point out that an in-depth survey and comparison of existing solutions to IoT security threats is not the ultimate objective of this paper. Instead, we aim to encourage research efforts to lay down fundamental basis for the development of new advanced security systems, algorithms, and methodologies for next-generation IoT systems.

## II. WHY IS IOT SECURITY SO CHALLENGING?

The notion of Internet of Things (IoT) has been used for the first time by Kevin Ashton of Procter & Gamble, while working on a slide show in 1999 [18]. Since then, there have been numerous attempts to define (or summarize) what the term IoT encompasses. In a nutshell, we can define the IoT as the inter-networking of physical devices, vehicles (also referred to as "connected devices" and "smart devices"), buildings, and any other item possessing an electronic component. These objects, also called "*things*," leverage network connectivity to collect and exchange data beyond what was previously imaginable. As far as its economic impact is concerned, it has been estimated that the IoT has a total potential economic impact of $3.9 trillion to $11.1 trillion a year by 2025. That level of value would be equivalent to about 11% of the world economy [19].

One of the most disruptive features of the IoT is that it allows objects to be sensed or controlled remotely across existing or ad-hoc network infrastructure, creating a myriad of opportunities for more direct integration of the physical world into computer-based systems. This results in improved efficiency, accuracy and economic benefit in addition to reduced human intervention. Furthermore, when sensors and actuators become involved, the IoT becomes an instance of the more general class of *cyber-physical systems* [20].

*A. A Brief Introduction to the Internet of Things*

Although some network architectures for the IoT have already been proposed [21–23], there is no general consensus on the details of how the IoT should be organized. On the other hand, most of the relevant industrial and academic parties have agreed upon that the main components of the IoT will be roughly orchestrated into the following components.

Arguably, the core components of any IoT architecture are the *IoT devices* or *things*. The concept of *thing* is by definition and on purpose very generic in scope. It may include, among others, objects such as smart devices (e.g., tablet, smartphones), or tiny sensors embedded in a car, coffee-maker, pencil, chair, and almost any object of common use. The common aspect among things is that they should be able to (i) capture different kinds of sensor data, such as location, images, sound samples, accelerometer data, biometric data, and barometric pressure; (ii) somehow interact with the environment (i.e., actuator), if the application requires so; and (iii) be uniquely identifiable and reachable from the Internet. In a broader sense, a thing may be also seen as a human equipped with a computing device, i.e., a human sensor, as proposed in [24]. This allows the IoT to gather more complex information about a phenomenon, for example, traffic status or weather information. A myriad of IoT devices has been already commercialized to implement extremely diverse IoT applications; for an excellent overview of these devices, the reader may refer to [25].

Once the sensed information has been collected, it becomes necessary to interact with the IoT devices through an *IoT network*. Similar to the concept of IoT thing, it is yet unclear how the IoT network will exactly function. In principle, the IoT network should be at least able to (i) handle the traffic generated by billions of devices in a scalable and reliable way; and (ii) adapt to the bandwidth, delay, and throughput requirements of different IoT applications. Some network protocols and standards have already been embraced by the research and industry community. For example, the IEEE 802.15.4, Bluetooth, WiFi, near-field communications (NFC), radio-frequency identification (RFID) are already part of almost any IoT device [14]. Besides these established technologies, the novel field of low-power wide-area networking (LPWAN), where IoT things can communicate at a very low bit rate for distances up to 30 kilometers in rural areas [26], is growing at an exciting pace. Unluckily, how these extremely diverse technologies should interact with each other has not been agreed upon. A comprehensive classification of existing technologies for networking in the IoT is beyond the scope of this paper; for excellent insights on the topic, we refer the reader to [27–29].

The back-end component of the system is the *IoT platform*, which is a suite of software components enabling the monitoring, management, and control of IoT devices. The IoT platform (i) facilitates remote data collection and integration with third party systems; (ii) exists independently between the hardware and the application layers of the IoT technology stack; and (iii) enables the implementation of IoT features and functions into any device in the same way. It is also responsible for the filtering, elaboration, and redistribution of sensed data. This component is usually implemented by a set of servers dedicated to the processing and storing of the sensed data, usually in relational databases, or databases specially adapted to the management of sensor readings. Forecasting the IoT explosion, the most established technology companies in the United States and abroad have proposed their own IoT platform, from Samsung Artik Cloud to Amazon AWS IoT, Google Cloud IoT and Microsoft Azure IoT suite, just to name a few [30].

*B. Unique Challenges in IoT Security*

Although the IoT presents features that are already present in other computer networking paradigms, we strongly believe that the IoT presents a completely different scenario and thus novel research challenges, especially as far as the security field is concerned. We believe the following point summarize the main reasons that should spur novel and transformative IoT security research in the near future.

– *Network and device size.* Handling the sheer size of the IoT will be a major problem from a security standpoint, since existing security protocols and technologies were not designed to scale to tens of billions of devices [31]. Furthermore, the stringent budget constraints of IoT manufacturers impose limited memory and computational power in IoT platforms, as well as the employment of tiny, inexpensive batteries for energy storage. Most importantly, since battery replacement will be extremely difficult or impossible (for example, when sensors are deployed on top of streetlight poles [32] or inside the human body [33]), such process will be tremendously expensive and time consuming. Thus, optimizing energy consumption becomes fundamental. In other words, the sheer number of devices coupled with the limitations in computation, memory and energy capabilities [25], strongly motivate the need for the design and adoption of novel security mechanisms capable of providing their functionalities without imposing excessive computational or storage burden on the IoT devices but also designed to be highly scalable.

– *Human component.* Among others, seamless human-machine interaction will be one of the most disruptive features of the IoT. Recent advances in machine learning and artificial intelligence will enable the IoT to learn and dynamically support our preferences and lifestyles at home, at work and on the move. Minuscule sensors will be able to perfectly deliver drugs [34] and capture biometric information [35] remotely, and provide doctors with a detailed view of our health condition. In other words, the information exchange will be mutual and intertwined – the IoT will learn from us and in turn, we will learn from the IoT. On the other hand, sharing information about ourselves, our home or business premise may become a liability in case that such information is accessed by malicious users or unwanted third parties. Thus, privacy and access control become a fundamental aspect in IoT. Also, another issue lies in that humans become primary actors of the sensing process in the IoT. On the other hand, there is no guarantee that humans will generate information reliably, for example, because they are not willing or not able to [36]. To address this paramount issue, novel *trust and reputation* mechanisms [10] will be needed, that will need to scale to billions of people (current human population).

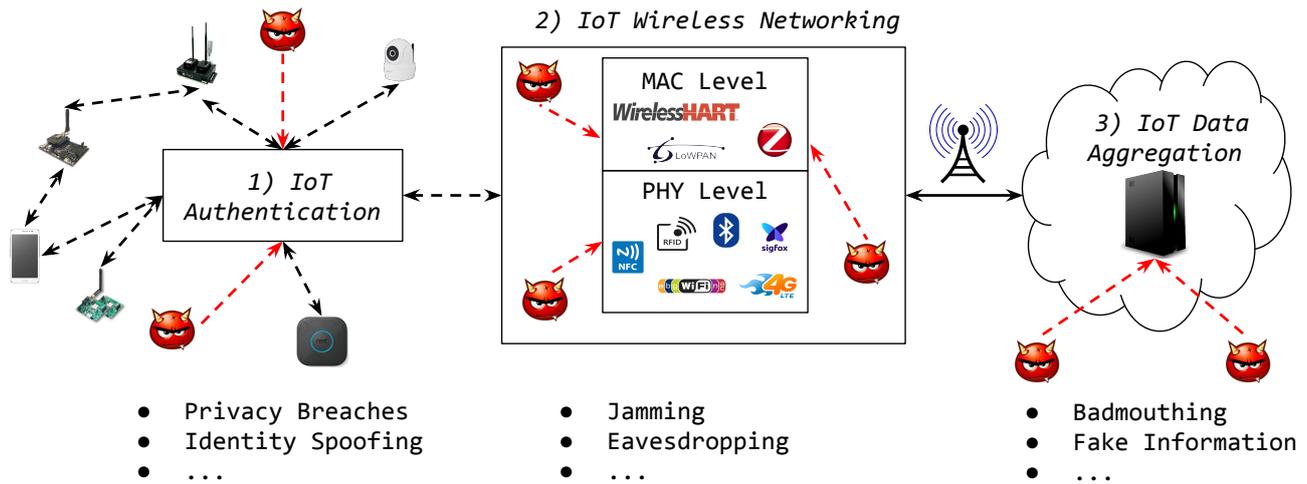

Figure 1. A "system" overview of security threats in the Internet of Things.

– *Heterogeneity*. The IoT is a complex ecosystem that will interconnect smartphones, tablets, machine type devices (MTDs), people, and mundane objects into a large-scale interconnected network. Because of this wide range of devices, a plethora of different IoT protocols, algorithms, and standards, will necessarily coexist, especially in the networking domain. While some manufacturers are adopting more open IoT standards such as MQTT [37] and the Internet Engineering Task Force (IETF) protocol stack for constrained IoT devices [38], nowadays most of the IoT is based on legacy systems that rely on proprietary technology, which has ultimately led to an anti-paradigm called the "Intranet of Things" [39]. Moreover, most of the existing research has made the assumption that there is a static association between the resources of the IoT and the surrounding real-world entities. On the contrary, the IoT environment is extremely heterogeneous and dynamic as IoT devices may experience unpredictable mobility, which results in sudden variations of communication capabilities and position over time [40]. Such an environment makes the resolution of available IoT devices a challenging task.

## III. TAXONOMY OF IOT SECURITY THREATS

While most of the IoT security threats are going to be inherited from the current Internet, a number of new threats that have yet to be discovered will also necessarily be unveiled in the near future. On the other hand, we can expect that the main attack strategies will be organized as illustrated in Figure 1, which also depicts the IoT scenario under consideration.

As discussed previously, the main target of an IoT system is to collect data from IoT devices deployed in strategic places, and if required, react to such information by controlling the environment accordingly (e.g., through actuators). We divide the data collection process in three steps, namely *i) IoT authentication, ii) IoT wireless networking,* and *iii) IoT data aggregation and validation.*

### A. IoT Device Identification and Authentication

Before being able to connect and exchange data with the IoT system, an IoT device must first be *identified*, and *authenticate* itself to prove that it is entitled to join the IoT system and transmit/receive data [41, 42].

Regarding identification, in [43] the authors present *IoT Sentinel*, a system capable of identifying the types of devices introduced to an IoT network and enforcing mitigation measures for device-types that have potential security vulnerabilities. The system does so by controlling the traffic flows of vulnerable devices, thus protecting other devices in the network from threats and preventing data leakage. Zenger *et al.* [44] proposed a vicinity-based pairing mechanism that delegates trust from one node to another based on physical proximity. Messaging between devices can be authenticated using multiple communication channels (e.g. Bluetooth + NFC) to ensure secure pairing [45]. Another approach to device identification is *device fingerprinting*, which leverages imperfections of hardware components (i.e., clock skew, RF signature, phase noise, and so on) to uniquely identify different wireless devices [46].

Given its paramount role toward an effective and efficient IoT, research on IoT authentication has gained a lot of traction in the research community over the last years. Traditionally, authentication has been mostly implemented by leveraging public-key cryptography (PKC). On the other hand, most of PKC authentication schemes are not amenable to be used in IoT scenarios, as PKC relies on computation-expensive modular exponentiation. For this reason, a significant number of authentication schemes based on lightweight encryption have been proposed. For an exhaustive survey on the topic, the reader may refer to [47].

Elliptic curve cryptography (ECC) [48] has been proposed as a valid alternative to PKC to address the constrained nature of IoT devices. ECC-based schemes have better performance thanks to the smaller key size – 160-bit ECC achieves the same security level as 1,024-bit RSA. However, ECC-based ones also require a certification authority (CA) to maintain a pool of certificates for users' public keys, and the users need extra computation to verify the certificates of others. Among recent work based on ECC, in [49] the authors present an authentication and access control method which establishes the session key on the basis of ECC. This scheme defines access control policies that are managed by an attribute authority, enhancing mutual authentication among the user and the sensor nodes, as well as solving the resource-constrained issue at application level in IoT.

A number of IoT-tailored authentication schemes using alternative encryption methods have also been proposed [50]. For example, an authentication protocol for IoT is presented in [51], using an encryption method based on XOR manipulation. A user authentication and key agreement scheme also based on XOR computations was proposed in [52] and then improved by [53], which enables a remote user to securely negotiate a session key with an IoT device, using a lean key agreement protocol. Baskar *et al.* recently proposed a lightweight encryption algorithm that uses a chaos map-based key [54], which was also implemented on a Field Programmable Gate Array (FPGA), Its performance is also compared with other lightweight algorithms in the literature. The algorithm is capable of achieving maximum throughput of 200 kbit/s on an FPGA with 1550 logic gates and 128 bit of key size.

Although the design of bullet-proof authentication mechanisms is always desirable, it generally implies high computational cost and/or trusted entities, a prerequisite that may be not applicable to several scenarios. For this reason, the Internet Engineering Task Force (IETF) has developed the Constrained Application Protocol (CoAP) [55], a lightweight scheme specifically tailored for energy- and resource-constrained IoT applications that relies on a dedicated security layer. This security layer, namely Datagram Transport Layer Security (DTLS), is built on top of the UDP protocol and provides IoT devices with several security features, together with a reliable authentication framework [56]. Authentication is usually achieved through handshake messages. For example, DTLS is leveraged in [57] to attain two-way authentication. In [58] the authentication process of DTLS is delegated to an external server that considerably offloads the network by reducing the overhead due to the handshake process. To provide authentication at the Internet layer, especially in the context of *6LoWPAN* systems, IPsec [59] has been identified as a suitable security framework [60]. Despite the fact that these works leverage IPsec, it has been shown that the energy consumption caused by the authentication mechanism envisioned by IPsec causes non-negligible energy consumption [61], which might not satisfy the energy-efficiency constraints required of most IoT applications.

The approaches discussed above are mainly tailored to provide authentication at the higher levels of the protocol stack. On the other hand, a promising and effective approach relies on the lower physical layer to provide authentication through RF fingerprinting [62]. Similar to human fingerprints, IoT devices are characterized by peculiar and unique features such as rising time of the RF signal and non-linearities. These peculiarities can be exploited to unequivocally identify a single transmitting device with high probability [62, 63]. The feasibility of RF fingerprinting has been demonstrated in [64], where a fingerprinting-based authentication system has been implemented on Raspberry Pi boards and cloud services. Though authentication of IoT devices is always desirable, there are some relevant applications of the IoT paradigm, such as medical and vehicular networks, where authentication is vital to guarantee the reliability of the network [65]. Both applications involve the transmission of sensitive data such as patients heart-rate and blood pressure [65, 66], or vehicle speed and traffic-related data [67].

A malicious node might be able to access the IoT network and generate fictitious and potentially dangerous messages. As an example, malicious users can harm patients by sending fake medicine release messages to IoT devices [66], or can trigger the automatic braking system by sending fake accident alerts to smart-vehicles [68]. It is clear that those scenarios require a highly reliable and secure authentication mechanism. A solution for authentication in health-care IoT scenarios is proposed in [69], where RF fingerprinting is employed together with biometric information to provide effective authentication in health-care IoT applications. Machine learning is used in [70] to design a risk-based authentication scheme that adapts the authentication protocol according to different risk levels. Similarly, a variety of authentication mechanisms for vehicular networks have been proposed in the literature by using electronic fingerprints of on-board devices [68].

Finally, some research efforts have focused on the design of authentication algorithms that mutually verify the identity of IoT devices. In fact, an imposter malicious node can aim to get access to the IoT network to eavesdrop ongoing communications and/or to take control over both the IoT devices and the IoT platform. Although desirable, mutual authentication comes at a cost in terms of overhead and computational complexity, which generally result in higher energy consumption. For this reason, much effort has been devoted to the design of lightweight and effective algorithms, such as [71–75]. For example, in [74] the authors exploits CoAP to design a mutual payload-based authentication scheme which replaces the native DTLS security framework with a lower-complexity authentication mechanism. Instead, a scalable and low-complexity approach is proposed in [75] where imperfect shared keys are considered.

### B. IoT Wireless Networking

Once the identity of the IoT device has been established, it becomes necessary to guarantee effective and efficient wireless communication between (i) the IoT devices themselves, if required; and (ii) the IoT devices and the IoT platform located in the cloud. To this end, depending on the particular application and the desired communication range, different wireless technologies may be used. Since TCP/IP will be the primary network/transport layers used in the IoT, researchers have been focused on addressing threats at the medium access control (MAC) and physical (PHY) layers [12, 76].

Since wireless transmissions are for the most part broadcast in nature, they are extremely vulnerable to eavesdropping [77], denial-of-service (DoS) [78], and jamming [13], among others. Although countermeasures for such attacks have been proposed, they are tailored to address only specific behavior of attackers [79] and specific wireless technologies [80], or employ cryptography solutions not applicable in the resource-constrained environment of the IoT.

Security of IoT wireless networking presents unprecedented challenges. This is because the IoT will use a plethora of different wireless technologies, *e.g.*, near-field, intra-body communications, *WiFi, Bluetooth* and *ZigBee* [81]. In addition, long-range technologies such as *Sigfox* [82] and *LoRa* [83] are expected to become major players in the near future [26]. The unprecedented diversity in the IoT networking environment calls for security solutions that are applicable to any IoT context and are flexible enough to deal with dynamic attack strategies and different technologies.

One of the most disrupting attacks to wireless communications is jamming, which is capable of rendering infeasible all ongoing communications within the attacked area. Thus, anti-jamming countermeasures for IoT applications have been widely investigated in literature [84]. Given the distributed nature of the network, game theory has been identified as a well-suited mathematical framework to provide effective anti-jamming mechanisms [85], where IoT devices act as the players of a game aiming at either avoiding or mitigating the disruptive attacks of the jammer. Other approaches rely on traditional jamming-proof mechanisms such as relaying [86, 87], spread-spectrum [88] and frequency hopping technologies [89].

Alongside jamming attacks, malicious entities may use eavesdropping to obtain sensitive and private information by monitoring ongoing transmissions between IoT nodes. Such an attack has been shown to be effective not only against traditional wireless transmission techniques such as WiFi and Bluetooth, but also against transmissions based on visible light communications (VLC) [90], which will be key in smart home applications. The most common approach to counterattack eavesdropping consists in the encryption of transmitted data [91] or the use of temporary pseudonyms or nicknames [92]. However, it is worth mentioning that other approaches have been proposed. For example, the use of relay nodes allows to generate new routing paths and bypass the eavesdropper. Although jamming is generally associated to disruptive attacks by malicious users, there are different cases where *friendly jamming* can, instead, be profitably exploited to improve the security level of the ongoing communications. Specifically, it is possible to take advantage of friendly jamming to reduce the signal-to-noise ratio level at the malicious node, so that an eavesdropper may not be able to decode transmitted data [93, 94].

A major concern in IoT networks consists in their vulnerabilities with respect to Distributed Denial-of-Service (DDoS) attacks. Indeed, the most disruptive attacks to the Internet have been carried out by exploiting IoT devices and their security breaches to generate bot-nets which consequently start DDoS flooding attacks [95]. It has been also shown [96] that a single compromised light bulb connected to the Internet can start a chain reaction and turn ON/OFF all the light bulbs of an IoT-enabled smart city. From the above discussions, it is straightforward to conclude that proper solutions are needed to avoid the generation of IoT-based bot-nets, and to protect the IoT system against attacks which aim at modifying the behavior of the whole network. Accordingly, several solutions have been proposed to either detect ongoing DDoS attacks [97], and to prevent them [98].

Other security threats which are worth mentioning are the replay and Sybil attacks [99]. Replay attacks are aimed at generating congestion and collisions in the network by eavesdropping and re-transmitting several copies of the same packet To avoid this attack, timestamps [100] may be employed. However, such an approach might be not completely secure if the attacker is capable of generating legitimate timestamps. Sybil attacks are instead targeted at generating fake virtual IoT devices which transmit synthetic packets through the network [101]. By means of Sybil attacks, malicious users can generate and transmit fake or corrupted measurements and messages to subvert the proper functioning of the network. To avoid such a threatening attack, intrusion detection mechanism [102, 103] have been proposed.

*C. IoT Data Aggregation*

Once the data generated by the IoT devices has been collected, it becomes necessary to infer *knowledge* based on such data. On the other hand, the information sent by two or more IoT devices about the same event is likely to be conflicting, for example, due to a noisy environment, malfunctioning, or network delay, among others. Furthermore, malicious or compromised IoT devices may also generate on purpose fake information and feed it to the system, to compromise the application's functionality at the information level. This becomes an extremely important issue when humans are directly contributing to the data collection process through their smart devices, such as in participatory sensing [104]. For example, in March 2014 students from Technion-Israel Institute of Technology successfully simulated through GPS spoofing a traffic jam on the *Waze* mobile application. The attack lasted for hours, causing thousands of motorists to deviate from their routes [105]. Similar Sybil attacks have also been successfully studied in the United States [106].

Over the years, the concept of *trust* and *truth* have been successfully employed in the computing, communication and networking fields [107]. In this vision, each network component is seen as a human, whereas the network itself may be seen as a society where decisions are made based on inputs from each node. This idea is often referred to as *technological trust*, a quantitative measure of trustworthiness for interactions between entities [108]. In the context of the IoT, applications are offered based on the information provided by the IoT devices, either through mobile devices or automated sensors. Therefore, the trustworthiness of the overall contributions from IoT devices is fundamental to guarantee the Quality of Information (QoI) of the IoT system itself.

For this reason, the fields of *trust estimation* and *truth discovery* have recently attracted enormous attention from the data mining and crowdsourcing communities [109, 110], due to its ability to estimate the reliability degrees of participants and thus infer correct knowledge based on the data without any supervision. On the other hand, existing algorithms were not designed to deal with the unprecedented number of devices that are expected to populate the IoT network. Specifically, it has been shown that the above approaches do not well-scale with the number of devices [111], and cannot directly be applied to the IoT environment. Accordingly, the research community has recently devoted significant effort on the design and assessment of algorithms that enjoy both low computational complexity and scalability.

The original concept of trustworthiness has its origins in human society. According to the Merriam-Webster dictionary [112], trust is the *"[...] belief that someone or something is reliable, good, honest, effective, etc."* In other words, trust is a qualitative way of expressing whether a particular interaction with an entity is going to be dependable, given evidence of outcomes of prior or current behaviors and interactions. Since IoT devices have limited interaction capabilities, estimating their trustworthiness is significantly challenging. Among others, fuzzy logic has been identified as an effective mechanism to evaluate the trustworthiness of IoT devices. By exploiting fuzzy logic sets, [113] has shown that it is feasible to design scalable trust estimation mechanisms for IoT applications. A similar approach is presented in [114], where authors enforce cooperation among things to develop a fuzzy-based reputation system that effectively allows to evaluate trustworthiness of IoT devices. Human relationships are leveraged in [115] to design dynamic trust estimation mechanisms which account for social affinities.

*Truth discovery* is generally related to trustworthiness, as trusted devices are expected to generate reliable and accurate information. However, due to the exorbitant number of interconnected IoT devices, scalable trust discovery processes are required as traditional solutions for other networking applications are expected to fail [111]. For example, [111] leverages trustworthiness and clustering algorithms to provide a trust discovery mechanism that also relies on problem scale reduction to achieve scalability. However, as outlined in [116–118] truth might not be unique and multiple truth could exist. Accordingly, proper truth discovery mechanism must be designed by exploiting, for example, Bayesian approaches [116] or predictions and implications [117]. Given the huge amount of data generated by the IoT network, data mining and machine learning approaches can be proficiently leveraged to provide improved services to citizens such as traffic management and e-health monitoring [119].

## IV. Research Challenges

Although the topic of IoT security is gaining increasing traction in the networking community, we believe that some important research challenges related to the field still remain substantially unexplored. We summarize and discuss the challenges below, hoping that the following roadmap will eventually stimulate discussion and further research and development in the IoT research community at large.

### A. Toward Secure-by-design IoT Systems

The most important takeaway from the previous discussion is that, as yet, IoT security has been so far approached in an *ad-hoc fashion*, where countermeasures are not planned for beforehand but instead temporary measures (i.e., "patches") are put in place when an attack is discovered. Considering that the IoT will work at a scale in the order of billions of devices, these patches are not adequate to address the need of homogeneous, standard, widely-adopted security procedures.

Our vision is the following. We believe that the complexity and the scale of the IoT require the enactment of a novel, holistic approach to IoT security, where security is approached in a *proactive* fashion and threats are addressed in a *scalable* and *reliable* manner. The IoT technology landscape of today is too complex and disruptive for security to be more than a set of loosely-integrated solutions. On the contrary, security must be deeply embedded in every stage of the production cycle, from product design to development and deployment. Too often, security tends to be an afterthought in development, and while there are exceptions, in many cases economic drivers or lack of awareness of the risks cause businesses to push IoT devices to market with little regard for their security. For these reasons, the concept of *security-by-design* [120] should be a main driving force in future IoT security research.

Security-by-design is an approach that has been traditionally applied to software and hardware development [121–123]. It seeks to make systems as free of vulnerabilities and impervious to attack as possible before the system is actually released to the market. This is usually achieved through measures such as extensive testing and adoption of best practices in programming. The security-by-design model contrasts with less rigorous approaches including security-through-obscurity, security-through-minority and security-through-obsolescence. Recently, world-renowned companies such as VMWare and Cisco have started embedding security-by-design concepts in their products [124, 125]. For example, VMWare proposes a system called *AppDefense*, which is based on the concept that chasing observed bad actions is not fruitful in the long run. Specifically, it delivers an intent-based security model that focuses on learning what the applications should do – the "known good" – rather than what the attackers do – the "known bad". Similarly, Cisco's *Identity Services Engine* uses a policy-based approach to simplify security, reduce risk, and provide end-to-end security.

Conversely from other technological fields, achieving security-by-design in the IoT is significantly challenging, given the network scale and heterogeneity of the IoT devices. For this reason, we need a practical yet comprehensive and effective framework that will help driving the adoption of security-by-design principles in the fast-paced, ever-changing IoT landscape. To this purpose, we propose a novel framework where security is seen as a control problem of an *IoT dynamic* *system*. In short, the IoT dynamic system will possess several *inputs* and a single *output*, which is the information generated by the IoT application and used to infer knowledge about events of interest. The output of the system is also used as feedback to "control" the IoT system and the environment itself, through the implementation of security measures (*mitigation module*) that are guided by a security analysis (*detection module*). Figure 2 illustrates a block diagram of the components of our envisioned IoT security framework. In this scenario, inputs may be both *legitimate* and *malicious* (i.e., security attacks conducted by ill-intentioned entities).

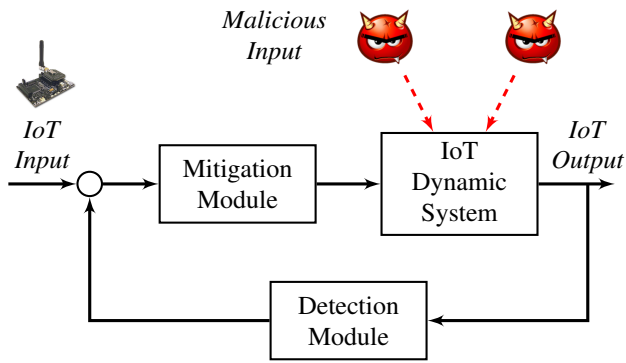

Figure 2. The IoT Dynamic System Control Problem.

The secure-by-design approach to IoT security offers several advantages with respect to previous paradigms. First, it provides a framework that abstracts from the specific security threat and tackles *classes* of problems, rather than a series of specific threats. Second, it stimulates IoT system designers to be *proactive* in considering security, and to come up with a security design plan that formalizes and addresses threats before their device/technology is released on the market. Third, it is flexible and scalable, since the control and learning modules can be designed and implemented both at the device and the system level to address different security threats, as we will discuss later.

Although the secure-by-design approach provides advantages, it also comes with novel and exciting technical challenges. We recognize two main research challenges toward the implementation of our IoT security framework, which are summarized below.

1) *Learning to Detect and Mitigate IoT Security Threats.* In every control and learning problem, the inputs and the state of the system must be properly modeled and formalized. This aspect is significantly challenging in the context of the IoT, as devices may generate significantly heterogeneous data (i.e., multimedia, text, sensory). After modeling states and inputs, it is necessary to design mechanisms able to detect and mitigate threats based on the current state and input. Furthermore, ML algorithms are *data-hungry*, meaning that they become more reliable as more and more data is used for training and testing [126]. Despite the importance of the issue, the research community still lacks large-scale datasets that collect information regarding attack activities. Although there have been efforts toward collecting crowd-sourced datasets in the wireless and mobile networking community, for example, as part of the Crawdad project [127], we need to collect and label data describing various IoT attacks as discussed in Section III.

2) *Design of Polymorphic Hardware and Software Modules to Enact Mitigation.* When a threat has been detected, it is necessary to enact countermeasures so as to swiftly mitigate the effect of the ongoing attack. This critical aspect requires the design of hardware and software modules able to "polymorphically" adapt to different requirements and thus swiftly put in action the necessary counterattack strategies.

As we have already outlined in Section II, IoT systems are extremely heterogeneous and dynamic in nature. This aspect, joint with the hardly predictable behavior of malicious entities, hinders significantly the design and development of effective threat detection systems. In this paper, we advocate the use of machine learning (ML) [128] and software-define networking (SDN) [129] to overcome these issues and implement autonomous and adaptive detection mechanisms for our IoT security framework. We provide a brief survey of the existing work below.

*1) Existing Work on Machine Learning and Software-defined Networking for IoT Security:* Recently, *machine learning techniques* have exhibited unprecedented success in classification problems in areas such as speech recognition [130], spam detection [131], computer vision [132], fraud detection [133], and computer networks [128], among others. One of the main reasons behind machine learning's popularity is that it provides a general framework to solve very complex classification problems where a model of the phenomenon being classified is too complex to derive or too dynamic to be summarized in mathematical terms.

Although ML can be considered a mature field, few works have applied ML techniques to solve issues related to IoT security. Recently, Zhang *et al.* have proposed a framework [17] to detect and mitigate cross-layer wireless attacks based on the application of Bayesian learning. Specifically, the framework establishes a probabilistic relation between an hypothesis (i.e., the attack is likely taking place) and the supporting evidence (i.e., there are signs of attack activities). This allows to update the hypothesis dynamically when new evidence is available. Therefore, the more evidence is gathered, the more accurate is the resulting hypothesis. The authors demonstrate through experiments and simulations that even small-scale malicious activities can still be detected with high confidence, as long as enough evidence is accumulated.

Cañedo and Skjellum use neural networks to develop a learning mechanism to assess the validity of information generated by IoT devices [134]. Similarly, in [135] the authors propose ARTIS, an artificial immune system that leverages ML algorithms to develop an adaptive immune system. Among the various properties, such as error tolerance, distributed computing and self-monitoring, ARTIS also makes possible to develop adaptive applications whose code evolves and branches according to past observations and experiences. However, the evolved code might be wrong and potentially vulnerable to security attacks, thus ARTIS also provides a learning-based anomaly detection mechanism to identify anomalies in the evolved code and in the data traversing the system. ML is also used in [136, 137] to develop an intelligent system that automatically detects security threats. Recently, in [138] authors use machine learning to develop a vulnerability assessment mechanism to identify and classify IoT devices according to their trustworthiness.

The development of algorithms and protocols based on *software-defined networking* [139] has gained tremendous momentum in the networking research community over the last years. In software-defined networks, components that have been typically implemented in hardware (e.g. mixers, filters, amplifiers, network protocols, etc.) are implemented in software, to ensure fast reconfigurability and adaptation to critical channel conditions (*e.g.*, significant multipath, Doppler effect, or path loss). The main downside of pure software-based solutions is that they completely trade-off reconfigurability for efficiency. On the other hand, we have recently seen a tremendous rise of wireless platforms based on the *system-on-chip* (SoC) concept [140]. These SoC platforms allow the design and implementation of customized hardware on the field-programmable gate array (FPGA) portion of the platform to achieve better performance [141].

The *software-defined* paradigm has gained increasing traction in the IoT research community [142–144]. Indeed, software-defined solutions ease the implementation of network elements and functions that were traditionally implemented and/or executed in hardware. Such process has been shown to be more energy-efficient, robust, scalable and adaptable than traditional hardware-based implementations [143, 145]. Furthermore, it allows for fast and efficient control and management of the IoT ecosystem, and has also paved the road for polymorphic and reconfigurable IoT systems. It is worth noting that IoT systems have several critical features (e.g., density of things, heterogeneity) that make traditional SDN solutions tailored for internet and data-center environments inapplicable [146]. Accordingly, IoT-specific SDN solutions have been proposed in the literature. As an example, [147] proposes an SDN-based IoT management framework that provides data traffic scheduling and routing mechanisms based on both network calculus and genetic algorithms. As an example, reprogrammability of IoT devices has been demonstrated in [148] where RFID devices have been successfully used to develop a generic and reprogrammable sensing platform for IoT applications.

### B. Model of IoT Network Inputs and Attacks

In order to work properly, machine learning (ML) algorithms need to have a clear and consistent formalization of the inputs (i.e., the data), states (i.e., attack/not attack), and outputs (i.e., processed data). Specifically, a clear definition of the input is fundamental to apply the concept of *dimensionality reduction*, which represents the process of feature selection and feature extraction [149]. Feature selection approaches try to find a subset of the original variables (also called features or attributes). There are three strategies: the filter strategy (e.g. information gain), the wrapper strategy (e.g. search guided by accuracy), and the embedded strategy (features are selected to add or be removed while building the model based on the prediction errors). In some cases, data analysis such as regression or classification can be done in the reduced space more accurately than in the original space.

Another major challenge is the proper definition of what an attack is and how to represent it. In other words, can we formalize and characterize (i) the "good" IoT network state (i.e., normal functioning) versus (ii) the "bad" IoT network state (i.e., attack is taking place)? Efforts of this sort have been conducted in [17], where the authors leveraged Bayesian learning to establish a relation between the probability that an attack is taking place and the evidence supporting it. On the other hand, we need to further investigate how to formalize *classes* of attacks and their effect of the network state. Furthermore, if an attacker has access to training data, or knows how the ML algorithms are trained, the attacker could eventually rely on this knowledge to tune its attack accordingly. In most cases, it is reasonable to assume that such information would not be available, since (i) supervised ML algorithms would be trained before deployment of the IoT nodes; and (ii) unsupervised ML algorithms would rely on parameters that could be encoded on the hardware chip of the IoT devices. However, the impact of such kind of attacks should be investigated by further research.

### C. Explore the Use of Reinforcement Learning

An area that is yet to be explored is the opportunity of leveraging unsupervised learning to implement secure-by-design IoT systems. Specifically, reinforcement learning (RL), which is ML inspired by behaviourist psychology, deals with how *agents* ought to take *actions* in an environment so as to maximize a cumulative reward. The problem, due to its generality, is studied in many other disciplines, such as game theory, control theory, operations research, information theory, simulation-based optimization, multi-agent systems, swarm intelligence, statistics and genetic algorithms, among others [150]. The advantage of RL for the IoT context is that there is no need for training or datasets, as the nodes can learn by themselves what is the right strategy to achieve the maximum reward according to the current state.

When modeling a problem with the RL framework, the environment is typically formulated as a Markov decision process (MDP). The main difference between the techniques based on dynamic programming and RL algorithms is that the latter do not need knowledge about the MDP and they target large MDPs where exact methods become infeasible. Reinforcement learning differs from standard supervised learning in that correct input/output pairs are never presented, nor suboptimal actions explicitly corrected. Instead the focus is on *on-line performance*, which involves finding a balance between exploration (of uncharted territory) and exploitation (of current knowledge) [151].

There have been a few number of attempts to leverage RL to model wireless networking problems, especially in the context of cognitive radio networks [152–154]. One of the most widely used forms of RL is Q-Learning [155]. However, it is still unclear how these technologies work in the context of IoT.

This is because RL suffers from the state-space explosion problems, which may make most of the existing RL algorithms inapplicable to the IoT context if undealt with.

*D. Blockchain for Decentralized IoT Security*

Although being originally designed to store and validate cryptocurrency transactions, the blockchain has recently attracted much in the IoT networking research community to address scalability and security problems [156–162]. The blockchain technology relies on decentralized, and thus scalable, consensus mechanisms that check, verify and store transactions in the blockchain while guaranteeing protection against data tampering attacks [163]. Given the sheer amount of IoT devices, the blockchain undoubtedly offers a suitable solution to the above problems.

The potential of the blockchain for IoT applications has been shown in [162], where the blockchain is used to provide a privacy-preserving IoT commissioning platform. Specifically, in [162] the data generated by IoT devices is made accessible from external service providers, while at the same time guaranteeing the anonymity of IoT devices and generating revenues for the owners of the devices. Since the blockchain guarantees protection against data tampering, it can be effectively used to verify the integrity and validity of software. For example, in [164] the blockchain is used to validate the different firmware versions embedded on IoT devices. The blockchain has also been exploited in both smart homing [159] and industrial scenarios [161] to protect local IoT networks and regulate traffic through distributed authentication mechanisms. Those works, however, showed that the blockchain inevitably generates computational overhead due to the blockchain mining algorithms, which ultimately increases both energy consumption and processing delay.

Furthermore, one of the major weaknesses of the blockchain is the so-called *51% attack* [165]. If any malicious node possesses at least 51% of the overall computational power, it might be able to monopolize the consensus mechanism and corrupt the integrity and trustworthiness of the whole network. The above discussion clearly shows that, while the blockchain might represent a true game changer in IoT applications, it is still unclear how we can efficiently adapt its mechanisms to constrained and pervasive networks such as the IoT.

## V. CONCLUSIONS

The IoT is revolutionizing the world around us by empowering every device, object and person to be connected to the Internet. With such massive presence of interconnected *things* deployed all around us, and in some cases *inside* us, the IoT offers exciting yet significant security research challenges that need to be addressed in the upcoming years. In this paper, we have provided our novel perspective on the issue of IoT security, which is based on a unique mixture of the notions of security-by-design, polymorphism, and software-defined networking. We have categorized and summarized the relevant state-of-the-art research, and proposed a roadmap of future research issues. We hope that this work will inspire fellow researchers to investigate topics pertaining to IoT security and keep on the race for a more secure technological world.

## ACKNOWLEDGEMENTS

We are grateful to the anonymous reviewers for their helpful comments, which have helped us to significantly improve the manuscript quality. Research was sponsored by the Army Research Laboratory and was accomplished under Grant Number W911NF-17-1-0034. The views and conclusions contained in this document are those of the authors and should not be interpreted as representing the official policies, either expressed or implied, of the Army Research Laboratory or the U.S. Government. The U.S. Government is authorized to reproduce and distribute reprints for Government purposes notwithstanding any copyright notation herein.

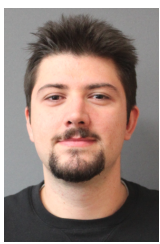

**Francesco Restuccia** (M'16) received his Ph.D. in Computer Science from Missouri S&T, Rolla, MO, USA in December 2016. Currently, he is an Associate Research Scientist with the Department of Electrical and Computer Engineering, Northeastern University, Boston, MA, USA. In the past, he has held research positions at the University of Texas at Arlington and the National Research Council of Italy. Dr. Restuccia's research interests lie in the modeling, analysis, and experimental evaluation of wireless networked systems, with applications to pervasive and mobile computing and the Internet of Things. He has served on the Technical Program Committee of IEEE INFOCOM (2018), IEEE WoWMoM (2017-2018), IEEE SMARTCOMP (2017) and IEEE LCN (2016-2018). He is a Member of the IEEE.

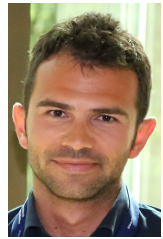

**Salvatore D'Oro** (S'12, M'17) received received his Ph.D. degree from the University of Catania in 2015. He is currently a Postdoctoral Researcher at Northeastern University. In 2015 and 2016, he organized the 1st and 2nd Workshops on COmpetitive and COoperative Approaches for 5G networks (COCOA), and served on the Technical Program Committee (TPC) of the CoCoNet8 workshop at IEEE ICC 2016. In 2013, he served on the TPC of the 20th European Wireless Conference (EW2014). Dr. D'Oro's research interests include game-theory, optimization, learning and their applications to telecommunication networks. He is a Member of the IEEE.

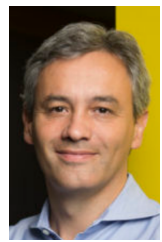

**Tommaso Melodia** (M'07, SM'16, F'18) received the Ph.D. degree in Electrical and Computer Engineering from the Georgia Institute of Technology in 2007. He is an Associate Professor with the Department of Electrical and Computer Engineering of Northeastern University. He is the Director of Research for the PAWR Project Office, a public-private partnership that is developing four city-scale platforms for advanced wireless research in the United States. Prof. Melodia's research focuses on modeling, optimization, and experimental evaluation of wireless networked systems, with applications to 5G Networks and Internet of Things, software-defined networking, and body area networks. Prof. Melodia is an Associate Editor for the IEEE Transactions on Wireless Communications, the IEEE Transactions on Mobile Computing, the IEEE Transactions on Biological, Molecular, and Multi-Scale Communications, Computer Networks, and Smart Health. Prof. Melodia's research is supported mostly by U.S. federal agencies including the National Science Foundation, the Air Force Research Laboratory, the Office of Naval Research, and the Army Research Laboratory. He is a Fellow of the IEEE and a Senior Member of the ACM.